# The Extragalactic Background Light (EBL): A Probe of Fundamental Physics and a Record of Structure Formation in the Universe

F. Krennrich & M. Orr
Physics and Astronomy Department, Iowa State University, Ames, IA 50011-3160
e-mail: krennrich@iastate.edu

**Physics Motivation:** Diffuse radiation fields permeate the Universe at all wavelengths. Encoded within these radiations is information regarding the fundamental components of the Universe and the physical mechanisms governing its formation and evolution. The Cosmic Microwave Background (CMB) is the most intense, and perhaps the most well known, of these radiative backgrounds and provides cosmological information about the very early Universe including the epoch of recombination when the Universe was ~300,000 years old.

The diffuse light spanning the UV/optical/near-IR/mid-IR/far-IR wavelength regime, termed the extragalactic background light (EBL), is second in intensity only to the CMB. The EBL is a source of gamma-ray opacity in the Universe, which provides a means for probing fundamental physics involving the nature of weakly interacting massive particles (WIMPs), and their role in early structure formation, as well as the existence of axion-like particles (ALPs), a class of particles arising within many grand unified theories. The EBL also possesses a wealth of information regarding processes associated with star/galaxy formation, particularly the collapses of massive stars, providing the dominant astrophysical contributions to the EBL (see Hauser & Dwek 2001 and Dwek & Krennrich 2013 for reviews).

In the context of particle physics, contributions to the EBL may come from the earliest stars whose fates were tied to a heavy dark matter particle. In the case of dark stars, their collapse would have been aided significantly by accretion of heavy dark matter particles; WIMPs in turn would have contributed to energy releases via WIMP annihilation and altered the properties of such massive stars significantly, prolonging their lifetimes, resulting in a noticeable (excess) contribution to the EBL in the mid-IR (Maurer et al. 2012). Recent EBL limits from γ-ray observations have ruled out dark stars with of order 100 solar masses, while more massive dark stars are still viable.

Another possible contribution to the EBL has been postulated to arise from the decay of exotic particles since the early universe (Bond et al. 1986). Such a contribution would result in a truly diffuse component, whereas radiation components of stellar/galaxy origin would be associated with point sources, that will eventually be resolved with current and/or next generation infrared telescopes.

Furthermore, the EBL offers a unique laboratory for fundamental physics through its ability to absorb γ rays via pair production. Particle physics processes that alleviate the γ-ray opacity of the universe have been proposed. These include the production of ALPs in the intergalactic magnetic field – the subsequent mixing of ALPs with photons within the γ-ray horizon of the observer could allow the detection of TeV photons from sources at redshift $z \geq 1$ (De Angelis et al. 2007; Sanchez-Conde et al. 2009), similarly to experiments shining laser beams through walls in the presence of a magnetic field.

**Methods for Studying the γ-Ray Opacity & EBL:** In order to separate the various components of the EBL, an accurate accounting of all EBL contributors is required. The key to such cosmic consistency tests involves the comparison of EBL constraints gained from different methods; a minimal EBL is set by the total amount of galaxy light present in the universe and is determined by galaxy counts with optical/IR telescopes. Upper limits to the EBL intensity are derived from assumptions regarding the maximum level of EBL absorption that can be present in γ-ray spectra of extragalactic sources ($\gamma + \gamma \rightarrow e^+ + e^-$).

Further reduction of the EBL limits from γ-ray measurements with CTA (Cherenkov Telescope Array; Actis et al. 2011), combined with deeper surveys of galaxy light (*James Webb Space Telescope* [JWST]) will result in either of the following outcomes:

1) The γ-ray constraints and galaxy count data converge. If substantial exotic truly diffuse contributions to the EBL can be ruled out, this EBL measurement will be an important pillar of cosmology, as the EBL from star formation is closely tied to the supernova rate and thus the flux of supernova neutrinos, and also depends on the X-ray background and number of dust enshrouded AGNs.

2) A residual diffuse component may be present if the γ-ray measurements demonstrate a higher EBL than is accounted for by galaxy counts. In this case tests, e.g., a spectral dependency of the diffuse component can be used to constrain a relic particle contribution.

3) The γ-ray measurements violate (i.e., intensities are lower than) the flux measurements from galaxy counts. In this case, the assumption of pair-production, i.e., $\gamma + \gamma \rightarrow e^+ + e^-$ in intergalactic space is not the only process affecting the propagation of γ-rays, and the opacity of the universe to TeV photons caused by the EBL is alleviated by additional processes, e.g., the mixing of axion-like particles (ALPs) with photons in the intergalactic magnetic field (see CF3: Non-WIMP Dark Matter).

In the latter two scenarios, measurements and constraints of the EBL obtained with CTA will provide important insight into physics both within and beyond the Standard Model of Particle Physics.

**Current and Planned Projects:** The Fermi-LAT has detected a redshift dependent absorption feature in the energy spectra of blazars and has led to a significant reduction in the uncertainties in the UV/optical-EBL. Several groups derived strong constraints in the near-IR to mid-IR using data from the MAGIC, H.E.S.S. and VERITAS atmospheric Cherenkov telescopes (Aharonian et al. 2006; Mazin & Raue 2007; Orr et al. 2011; Abramowski et al. 2013). Figure 1 provides an up-to-date summary of EBL measurements, limits, and constraints.

CTA will provide the necessary data to expand these measurements further into the mid- and far-IR wavelength regimes of the EBL, providing valuable constraints on the role of dark matter in the formation and evolution of the first stars in the early Universe. These observations will also yield measurements of γ-rays at energies (~100 GeV to 10 TeV) where EBL absorption is expected to be substantial (> ×10). It is over this energy range that the U.S. contribution of medium-sized telescopes for CTA will be most sensitive. These observations will provide a dramatically improved sensitivity, compared to the current generation of instruments, and probe for evidence of exotic physics such as ALP-photon mixing.

**Key Issues and Future Steps:** Helgason & Kashlinsky 2012 (see also Franceschini et al. 2008) have recently reconstructed the EBL using a direct observational approach completely independent of the theoretical modeling of galaxy population evolution. This is made possible through the use of deep galaxy surveys performed at UV through IR wavelengths. The authors use these observations to construct empirical galaxy luminosity functions (LFs). The resulting EBL spectral energy distribution (SED) reconstructed by Helgason & Kashlinsky (2012) in the mid-IR is in excess of the upper limits obtained by Orr et al. (2011), obtained using γ-ray observations of blazars (Figure 1). The work of Horns & Meyer (2012), using source spectra from atmospheric Cherenkov telescopes, also favors a low mid-IR intensity of the EBL. This mismatch between direct and indirect constraints/measurements is extremely interesting as it could be indicative of yet to be detected physical mechanisms such as ALP-photon mixing.

There are also indications of a mismatch between indirectly (γ-ray) and directly (observed LFs) obtained constraints/measurements of the EBL in the near-IR. In this case, the disagreement is reversed – in that the γ-ray constraints yield higher limits than those obtained from direct observations. In particular, the constraints obtained by Orr et al. (2011) and Abramowski et al. (2013) are in excess of (or perhaps *marginally* consistent with) the reconstructed EBL of Helgason & Kashlinksy (2012) (Figure 1). This leaves ample room for a diffuse component of the EBL that remains inaccessible to galaxy surveys but still affects the absorption of γ-rays and hence the constraints obtained through γ-ray observations. Such a diffuse component could be the result of decaying weakly interacting relics of the Big Bang (Bond et al. 1986; Wang & Field 1989; Sciama 1998).

Furthering studies of the EBL require several orthogonal approaches; one is to perform γ-ray observations of blazars with much improved sensitivity, energy resolution, and sky coverage (future CTA) as well as deeper galaxy surveys. The latter will be provided by JWST (having a scheduled launch date of 2018) while the former will require an instrument such as CTA (Mazin et al. 2012). The enhanced performance of CTA in all aspects (e.g., sensitivity, energy resolution, sky coverage, etc.) will increase the size of the available γ-ray blazar population by approximately one order of magnitude (Sol et al. 2013). This will facilitate the detection of an EBL absorption signature across optical through far-IR wavelengths – potentially allowing for the actual measurement of the EBL spectral energy distribution and providing a true measurement of the γ-ray opacity of the Universe. The increased blazar population size will also provide a significantly improved sampling of source redshifts, a critical component to investigating the evolution of the EBL over cosmic time. However, in the mean time high-redshift blazar surveys with current generation instruments, e.g., VERITAS (Holder et al. 2011), are capable of catching a glimpse of new physics; the detection of TeV photons from a flaring blazar at redshift $z > 1$ would constitute a major discovery, and either require new physics (axion-like particle mixing) or other mechanisms for photons ''tunneling'' through the EBL. The availability of Fermi, HAWC and VERITAS data over the next 5 years provides a unique opportunity through Fermi's all-sky coverage at GeV energies, HAWC's northern-sky monitoring capabilities in the TeV regime, and VERITAS' excellent sensitivity.

Studies of the EBL with VERITAS, HAWC and CTA will bridge the gap between γ-ray astronomy, optical/IR astronomy, and particle physics. This is an integral component to developing a coherent picture of the early Universe, the physical processes responsible for its evolution, and ultimately explaining the Universe as it is known today.

**Acknowledgements:** The authors are grateful to M. Persic for helful comments. FK acknowledges support from the U.S. Department of Energy Office of Science and from the NASA Fermi Guest Investigator grant NNX11AO38G.

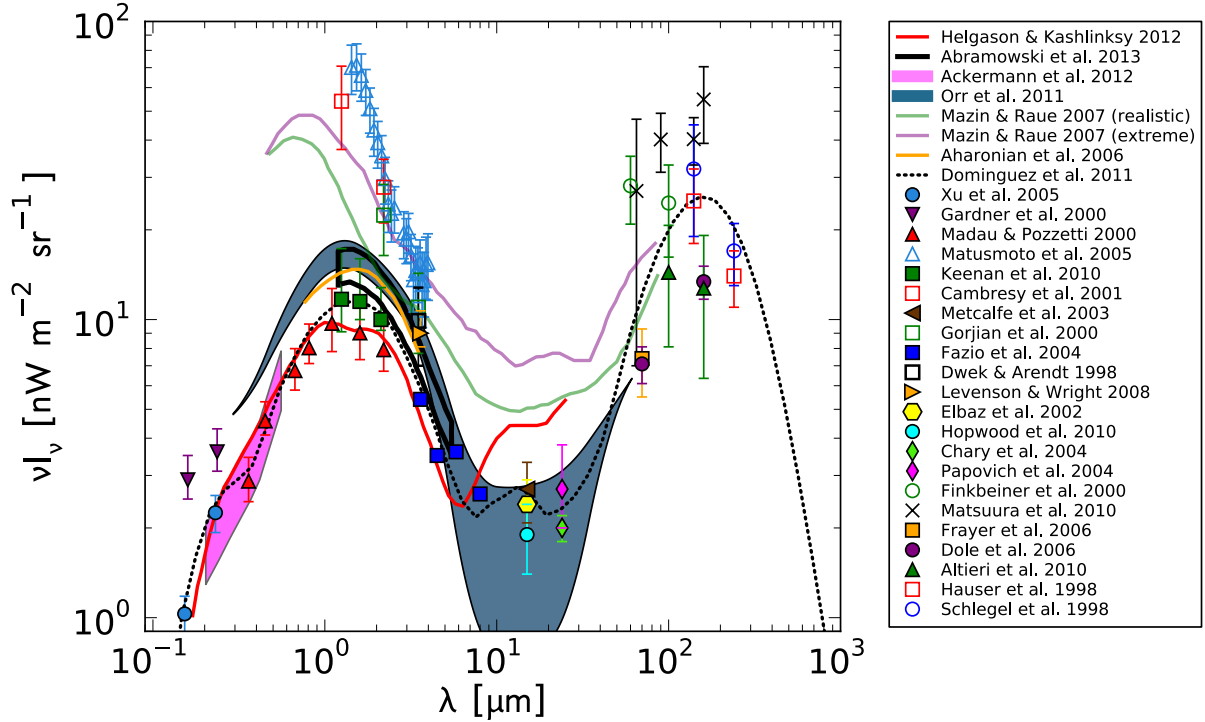

**Figure 1:** Summary of the EBL from direct measurements (open symbols), lower limits from galaxy counts (filled symbols), observed galaxy LFs (legend listing 1), constraints from IACT observations of blazars (legend listings 2-7), and the model of Dominguez et al. (2011) (legend listing 8). Other references can be found in Dwek & Krennrich 2013.